\documentclass[10pt,twocolumn,twoside]{IEEEtran}

\usepackage{graphicx}
\usepackage{multirow}
\usepackage{subfigure}
\usepackage{url}
\usepackage{epstopdf}
\usepackage{color}
\usepackage{xcolor}
\usepackage{bbm}
\usepackage[cmex10]{amsmath}
\usepackage{amssymb}
\usepackage{array}
\usepackage{dsfont}
\usepackage{caption}
\usepackage{multirow}
\usepackage{draftfigure}
\usepackage{url}
\colorlet{mygreen}{green!60!gray}

\begin{document}
%
\title{Primary Quantization Matrix Estimation of
Double Compressed JPEG Images via CNN}
%
%
%

\author{Yakun Niu, Benedetta Tondi,~\IEEEmembership{Member,~IEEE}, Yao Zhao, \IEEEmembership{Senior Member, IEEE}, Mauro Barni,~\IEEEmembership{Fellow,~IEEE} %
\thanks{This work was partially supported by the National Key Research and Development of China (No. 2016YFB0800404), the China Scholarship Council.}
\thanks{Y. Niu and Y. Zhao are with Institute of Information Science, Beijing Jiaotong University, Beijing, 100044, China, and also with Beijing Key Laboratory of Advanced Information Science and Network Technology, Beijing, 100044, China. (\{niuyakun and yzhao\}@bjtu.edu.cn).}
\thanks{M. Barni, B. Tondi are with Dept. of Information Engineering and Mathematical Sciences, University of Siena, Italy (barni@dii.unisi.it, benedettatondi@gmail.com).}

%
\vspace{-0.5cm}
}

\maketitle

\begin{abstract}
Available model-based techniques for the estimation of the primary quantization matrix in double-compressed JPEG images work only under specific conditions regarding the relationship between the first and second compression quality factors, and the alignment of the first and second JPEG compression grids. In this paper, we propose a single CNN-based estimation technique that can work under a very general range of settings. We do so, by adapting a dense CNN network to the problem at hand. Particular attention is paid to the choice of the loss function. Experimental results highlight several advantages of the new method, including: i) capability of working under very general conditions, ii) improved performance in terms of MSE and accuracy especially in the non-aligned case, iii) better spatial resolution due to the ability of providing good results also on small image patches.
\end{abstract}


\IEEEpeerreviewmaketitle


\section{Introduction}
\label{sec.intro}

%
%

Detection of double (or even multiple) JPEG compression plays a major role in image forensics since double compression reveals important information about the past history of an image \cite{Jessica2008,Li2008,Barni2017}.
Even more information can be inferred by estimating the primary quantization matrix used for the former compression. As an example, given an image with several  copy-pasted regions, it is possible to identify the different origin of the tampered areas by recognizing that they are characterized by different primary quantization matrices.

In some works, the primary quantization matrix is derived indirectly by estimating the primary  Quality Factor ($QF$)  \cite{Li2008,PQFEstimtion2011,Farid2009}. This, however, is a less general approach since the $QF$ is not a standard JPEG parameter and it may be undefined for compression packages using a proprietary quantization matrix (e.g. Photoshop).
Several model-based methods have been proposed for the estimation of the primary quantization matrix. Some of these methods exploit statistical modeling of DCT coefficients \cite{Bianchi2012,Cogranne2019}, while others reveal the presence of peculiar patterns induced by successive quantizations \cite{Lukas2003,Galvan2014}. A common feature of all these techniques is that they work only under particular conditions, e.g. they may require that $QF_{1} < QF_{2}$, or that the second JPEG compression is either aligned or non-aligned with the $8 \times 8$ grid of the first compression. For instance, the method in \cite{Galvan2014} works only when the two compressions are aligned and $QF_{1} < QF_{2}$. Similarly, \cite{Cogranne2019} is tailored for the aligned case, and it cannot estimate the first quantization step when this is a divisor of the second one. The system proposed in \cite{Bianchi2012} can work both in the aligned and non-aligned cases, however the performance when $QF_{1} > QF_{2}$ drop significantly. Eventually, the method presented in \cite{Dalmia2018} works in the non-aligned case only.
Another problem of approaches based on statistical analysis, is that their performance decrease significantly when they are applied to small patches. This is a significant drawback, especially when the estimation of the primary quantization matrix is used for tampering localization, since it reduces the minimum size of the tampered regions that can be identified in this way.

In this paper, we propose a method for primary quantization matrix estimation based on a Convolutional Neural Networks (CNN),  that is able to work under very general conditions, and on relatively small patches.
With regard to the network structure, we resort to a dense CNN architecture \cite{huang2017densely}, which we modified to tune it to the specific estimation problem considered here.
Similar architectures have been recently adopted for several image forensic applications, e.g. \cite{chen2019multi,kamal2018applicationDense,Huang2018Dense},
yielding, in most of the cases, improved performance compared to more traditional architecture (e.g. residual-based networks).
In particular, in \cite{Huang2018Dense} the authors successfully trained a dense model for detecting D-JPEG in the most difficult case where the same quantization matrix is used for both quantization steps.
We paid a special attention to the choice of the loss function used during training. In fact, we found that the most natural choice of defining the loss function as the squared Euclidean ($L_2$) distance between the estimated quantization steps and the true ones weighs too much outliers thus resulting in a lower accuracy of the estimation.

The experimental results we got demonstrate that the proposed approach greatly outperforms state-of-the-art techniques especially for small
patch sizes. The improvement is particularly significant when the second compression is not aligned to the first one, which corresponds to the most common situation in the case of copy-paste tampering, and when $QF_1 > QF_2$.
%
A significant advantage of the new method is that it works under very general conditions ($QF_1 < QF_2$, $QF_1 > QF_2$, aligned and non-aligned double compression).
This represents a big advantage over the state-of-the-art since a method capable to work reasonably well in all situations has never been proposed.



\section{Background and notation}
\label{sec.background}

Let $Q$ denote the $8 \times 8$ matrix with the quantization steps of the DCT coefficients, namely, the quantization matrix.
Double compression occurs when an image is first  JPEG compressed with a given  $Q_1$, decompressed (decompression involves de-quantization and inverse DCT), and then compressed again with a second quantization matrix $Q_2$.
%
%
We denote by ${\bf q}_1$ the 64-dim vector with the elements of $Q_1$, taken in zig-zag order \cite{pennebaker1992jpeg}.
It is known that the quantization steps corresponding to the medium-high frequencies are more difficult to estimate accurately, since they are quantized more heavily. However, they are usually less important in image forensics analysis, since they are not very discriminative given that they tend to be similar for most quantization matrices \cite{pennebaker1992jpeg}. For this reason, and following prior works \cite{Bianchi2012,Cogranne2019,Galvan2014}, we restrict the estimation to the first $N_c$ elements of ${\bf q}_1$.
%
In the following, we denote  with $({\bf q}_{1})_{N_c} = [q_{1,1}, q_{1,2},...,q_{1,N_c}]$  the vector of the first $N_c$ coefficients of ${\bf q}_1$.

The second compression can be either aligned or  non-aligned to the first one, depending on the position of the $8\times8$ JPEG compression grid. A misalignment occurs, for instance, when the image is cropped between the former and second compression stage, or in splicing cases, when a region of a single-JPEG image is copy-pasted into another image (in which case, very likely, the alignment between the compression grids is not preserved).

For convenience, in the rest of the paper,  we often refer to the JPEG Quality Factor ($QF$).
Whenever defined (the $QF$ is not a standard JPEG parameter), each $QF$ value identifies a specific quantization matrix $Q$.
Specifically, we denote with $QF_2$ the second compression $QF$ and with $QF_1$ the former.

\section{Proposed CNN-based $Q_1$ estimator}
\label{sec.proposedMethod}

%

%
%

As we said, our goal is to develop a method for $Q_1$ matrix estimation that can work under as general as possible working conditions, and also on relatively small patch sizes.
%
To accomplish this task, we designed an estimator that exploits the capability of deep learning architectures, where the information characterizing the former quantization steps is automatically learned from the image patches during training. In particular, we followed a recent trend in image forensics according to which no pre-processing is applied to the image under analysis before feeding it to the network \cite{chen2019multi,kamal2018applicationDense,Huang2018Dense}.

With regard to the network architecture, we modified the dense CNN architecture originally proposed in \cite{huang2017densely} (DenseNet) to tailor it to our estimation problem.
%
A peculiarity of the dense structure is that it connects each layer to every other layer in a feed-forward fashion, inside a dense block. Then, the features  extracted by early
layers are directly used by deep layers throughout the same
dense block, thus conferring to the learned features a hierarchical
structure.
It is argued in \cite{huang2017densely} that   the dense connectivity alleviates the vanishing gradient problem and strengthens the feature propagation;  moreover, it is possible to reduce the number of parameters of the trained model (instead, the number of links increases, passing, for each dense block, from $l$ to $l(l-1)/2$, where $l$ is the number of layers in the block).
%
Specifically, by referring to the original dense structure described in \cite{huang2017densely}, we considered a network depth of $40$, with  3 dense blocks and growth rate $k = 12$.
Each dense layer consists of 12 convolutional layers (all the convolutions have kernel size $3 \times 3 \times k$) and a transition layer, where $2\times 2$ average pooling is performed to reduce the input size.
The dropout is set to the default parameter $0.2$.
Before entering the first dense block, a convolution with $2 k$ $3 \times 3$ filters is performed.
Then, each convolutional layer $l$ has $k(l-1) + 2k$ input feature maps.
After the last dense blocks, global average pooling is performed, and the feature vector (of size $456$) is fed to the fully connected  layer.
The number of final output nodes is set  to $N_c$. We depart from a classification-like architecture, and no softmax is applied.


With regard to training, given the $N_c$ quantization steps that we want to estimate,  the CNN is trained to minimize the difference between the predicted values
and the true vector $({\bf q}_{1})_{N_c}$.
We tested two loss functions, namely the $L_2$ distance between the true quantization steps and the output of the network, and the log-cosh loss function, i.e. the logarithm of the hyperbolic cosine of the prediction error \cite{neuneier1998train}.
The rationale behind the use of the log-cosh loss is that it weighs less the presence of outliers favouring the reduction of the error on common examples. In this sense, the log-cosh has a behaviour similar to a loss function based on the $L_1$ distance, while being more easy to train
\footnote{With the $L_1$ loss, the gradient is always the same for large and small loss values; moreover, the derivative is not continuous, due to the discontinuity in 0. These two features make a CNN based on the $L_1$ loss difficult to train.}. More formally, we let:
%
%
\begin{equation}
\mathcal{L}({\bf x}) = \frac{1}{N_c} \sum_{i=1}^{N_c} \log(\cosh(q_{1,i}({\bf x}) - f_i({\bf x}))),
\end{equation}
where ${\bf f}({\bf x}) = [f_1({\bf x}),f_2({\bf x}),..., f_{N_c}({\bf x})]$ is the vector of the soft outputs, namely the predicted vector, and ${\bf x}$ the image patch under analysis.
Note that, given $t \in \mathds{R}$, $\log(\cosh(t))$ is approximately equal to $(t^2)/2$ for small $t$ and to $|t| - \log(2)$ for large $t$, meaning that the log-cosh works mostly like the $L_2$ norm
when the loss is small, but it is not affected strongly by an occasional widely incorrect prediction.
As we will show in Section \ref{sec.expRes}, the use of the log-cosh function provides better results in terms of prediction accuracy and hence we preferred it to the more classical $L_2$ loss.


The estimation provided at the output of the network is a real number. To get an integer estimation, the output of the network is rounded to the nearest integer. In this way the final prediction is obtained as $(\hat{{\bf q}}_{1})_{N_c} = \text{round}(\bf{f}({\bf x}))$,
where rounding is performed on each element of $\bf{f}$ independently \footnote{If the estimator knows in advance that a standard $Q$ is used, better results could be obtained by applying vector quantization. We did not do so, to avoid putting any restrictions on the $Q$ used for the first compression.}.
%
For a given image ${\bf x}$, the Mean Square Error (MSE) of the estimation is given by
%
%
${\text{MSE}}({\bf x}) = (1/N_c)(\sum_{i=1}^{N_c} |q_{1,i}({\bf x}) - \hat{q}_{1,i}({\bf x})|^2)$.
Notice that, due to rounding, any difference between the true and the predicted value smaller than 0.5, results in a $0$ error.
We also evaluate the accuracy of the estimation by means of the average prediction accuracy defined as follows: Acc$({\bf x}) = (1/{N_c})\sum_{i=1}^{N_c}\delta({q}_{1,i}({\bf x}),\hat{{q}}_{1,i}({\bf x}))$, where $\delta(t,v) = 1$ if $t = v$, $0$ otherwise.

\section{Experimental results}
\label{sec.exp}

\subsection{Methodology} 
\label{sec.method}

%

Since $QF_2$, or, more in general, the quantization matrix of the second JPEG compression,
can always be obtained from the JPEG file (e.g. read from the header), or very accurately estimated \cite{bestagini2012video,BarniIWBF18}, we trained the network by fixing $QF_2$. We verified experimentally that a network trained with a given $QF_2$, generalizes pretty well to different $QF_2$'s (hence, more in general, to a different $Q_2$ matrix), at least when the mismatch is not too strong.

To build the training and testing datasets, we started from color, uncompressed, never-processed images. The images are initially split into a training and a test set, then,
they are compressed first with several $QF_1$'s and then with the prescribed $QF_2$ (both $QF_1$ values larger and smaller than $QF_2$ are considered). A uniformly distributed random grid shift is applied between the two compressions to simulate JPEG misalignment. In this way, aligned compression occurs with probability 1/64.
The images are then cropped into patches of size $64\times 64\times 3$.
In order to avoid picking up too many patches from the same image and enforce diversity in the selected patches, we put a limit on the maximum number of patches from the same image, set to 100, and selected them from random image locations.
For testing, we considered 5 random patches per image.
%

\subsubsection{Settings}

We carried out the experiments on the RAISE dataset \cite{RAISE8K}  (for training and testing)  and also on the Dresden dataset \cite{Dresden} (for testing only).
4000 tiff images taken from RAISE were considered for training, for a total of $4\times 10^5$ patches per each $QF_1$ (additional 400 images were used for validation, contributing with $4\times 10^4$ patches).  5780 patches taken from 1156 tiff images were used for testing.
For the tests under mismatch conditions, all the 1488 images were taken from Dresden dataset (for a total of 7440 patches).
To build the training dataset, JPEG compression was performed with OpenCV.
For the testing dataset, in addition to OpenCV, we also used images compressed with Photoshop to carry out the first JPEG compression.
In the following, we provide results for the cases of $QF_2 = 90$ and $QF_2 = 80$.
The $QF_1$'s considered for training are respectively, $QF_1 = \{60,65,70,75,80,85,90,95,98\}$ for $QF_2 = 90$, and $QF_1= \{55,60, 65,70, 75, 80, 85, 90, 95\}$ for $QF_2 = 80$.
Finally, in all the experiments, we set $N_c = 15$,
which is the value considered in most prior works \cite{Bianchi2012,Cogranne2019,Galvan2014,Dalmia2018}.
%

The CNN has been implemented by using TensorFlow, via the Keras API, starting from the DenseNet implementation in \cite{kerasDense}.
The Adam optimizer was used with learning rate $10^{-5}$. 
The batch size for training and testing was set to 32 images. We got our first model for $QF_2 = 90$, by training the network for 60 epochs.
For $QF_2 = 80$, the models were trained for 20 epochs, by starting from the network trained on $QF_2 = 90$.

\begin{table*}[htbp] 
\scriptsize
\centering
\caption{ Performance of the CNN-based estimator (MSE/Acc), for $QF_2 = 90$,  in the various cases, on RAISE dataset. The comparison with the state-of-the-art is also reported.}
    \begin{tabular}{|p{0.4cm}<{\centering}|p{0.9cm}<{\centering}p{1cm}<{\centering}p{1cm}<{\centering}p{0.9cm}<{\centering}
 |p{1.5cm}<{\centering}p{1cm}<{\centering}|p{0.9cm}<{\centering}p{1cm}<{\centering}p{0.9cm}<{\centering}p{0.9cm}<{\centering}
 |p{1.2cm}<{\centering}p{1cm}<{\centering}|}
    \hline
    \multirow{2}{*}{$QF_1$}& \multicolumn{6}{c|}{Non-Aligned}& \multicolumn{6}{c|}{Aligned}  \\ \cline{2-13}
    & CNN-L$_{2}$& CNN-Log & \cite{Bianchi2012}, N-Al &  \cite{Dalmia2018}  &  \text{ \cite{Bianchi2012}, N-Al (128) } & \text{ \cite{Dalmia2018} (128) }& \text{CNN-L$_{2}$} & CNN-Log &  \text{ \cite{Bianchi2012}, Al  } & \text{ \cite{Galvan2014} } &  \text{ \cite{Bianchi2012}, Al (128) } & \text{ \cite{Galvan2014} (128) }\\
    \hline
  60&2.33/0.39&\textbf{2.17}/\textbf{0.54}&33.2/0.13&16.4/0.32&16.2/0.50&10.2/0.44&\textbf{1.86}/0.59&1.87/\textbf{0.68}&16.1/0.62
  &14.2/0.60&11.8/0.72&12.4/0.69 \\\hline
  65&1.41/0.51&\textbf{1.39}/\textbf{0.56}&25.6/0.12&16.9/0.30&12.8/0.49&9.49/0.45&\textbf{1.13}/\textbf{0.60}&1.28/0.57&11.5/0.67
  &9.21/0.62&8.43/0.76&9.53/0.72\\ \hline
  70&0.73/0.55&\textbf{0.73}/\textbf{0.62}&21.1/0.11&16.1/0.22&10.4/0.49&8.07/0.36&\textbf{0.56}/\textbf{0.71}&0.64/0.70&8.31/0.63
  &5.22/0.66&5.55/0.77&4.58/0.74 \\\hline
  75&0.46/0.65&\textbf{0.43}/\textbf{0.73}&21.9/0.09&20.7/0.20&12.6/0.39&10.2/0.35&\textbf{0.30}/0.77&0.31/\textbf{0.82}&17.3/0.50
  &3.16/0.70&5.66/0.78&2.72/0.79\\\hline
  80&\textbf{0.36}/0.76&0.39/\textbf{0.78}&26.6/0.07&27.6/0.08&19.0/0.19&13.4/0.22&\textbf{0.23}/0.85&0.27/\textbf{0.86}&15.9/0.36
  &1.61/0.73&4.95/0.69&1.56/0.80 \\\hline
  85&0.43/0.77&\textbf{0.38}/\textbf{0.81}&35.8/0.07&37.9/0.07&27.6/0.10&19.9/0.11&0.36/0.79&\textbf{0.30}/\textbf{0.84}&35.5/0.16
  &1.29/0.82&30.2/0.31&1.11/0.88\\\hline
  90&0.64/0.71&\textbf{0.63}/\textbf{0.72}&50.4/0.08&53.1/0.00&40.4/0.09&32.5/0.00&3.27/\textbf{0.05}&3.63/0.03&46.4/0.00
  &\textbf{2.33}/0.00&42.3/0.00&1.71/0.00 \\\hline
  95&1.06/0.72&\textbf{0.94}/\textbf{0.74}&70.0/0.05&75.3/0.00&56.9/0.04&48.9/0.00&0.75/\textbf{0.81}&\textbf{0.70}/0.79&66.1/0.13&
  8.72/0.00&62.7/0.17&8.19/0.00 \\\hline
  98&1.13/0.68&\textbf{0.90}/\textbf{0.77}&77.5/0.00&77.8/0.00&63.2/0.00&55.1/0.00&1.22/0.67&\textbf{0.96}/\textbf{0.77}&70.5/0.08
  &11.7/0.00&63.2/0.08&10.9/0.00 \\\hline
    \end{tabular}%
\label{tab.Res90}%
\end{table*}%

\subsection{Results}
\label{sec.expRes}
%

%

The performance for the case $QF_2 = 90$  are shown in Table \ref{tab.Res90}.
The comparison with the other methods is reported for the same 15 coefficients.
We see that the CNN-based estimator yields a significant improvement, especially in terms of MSE, which is much lower compared to state-of-the-art (sota) methods. For sake of completeness, we also report the results obtained by a CNN trained with an $L_2$ loss, which as anticipated, gives slightly worse performance in terms of accuracy than those obtained with the log-cosh loss.
In all the cases, the improvement with respect to the sota is evident.
In particular, the CNN estimator greatly outperforms all the other methods when $QF_1 > QF_2$, which corresponds to the most difficult scenario. The only case where results are not satisfactory (as with sota methods) is when $QF_1 = QF_2$, a case for which ad-hoc solutions must be adopted \cite{yang2014effective}
%
\begin{figure}[htbp]
\centering
\subfigure{
\label{Fig1_1}
\includegraphics[width = 0.65\columnwidth, width = 0.48\columnwidth]{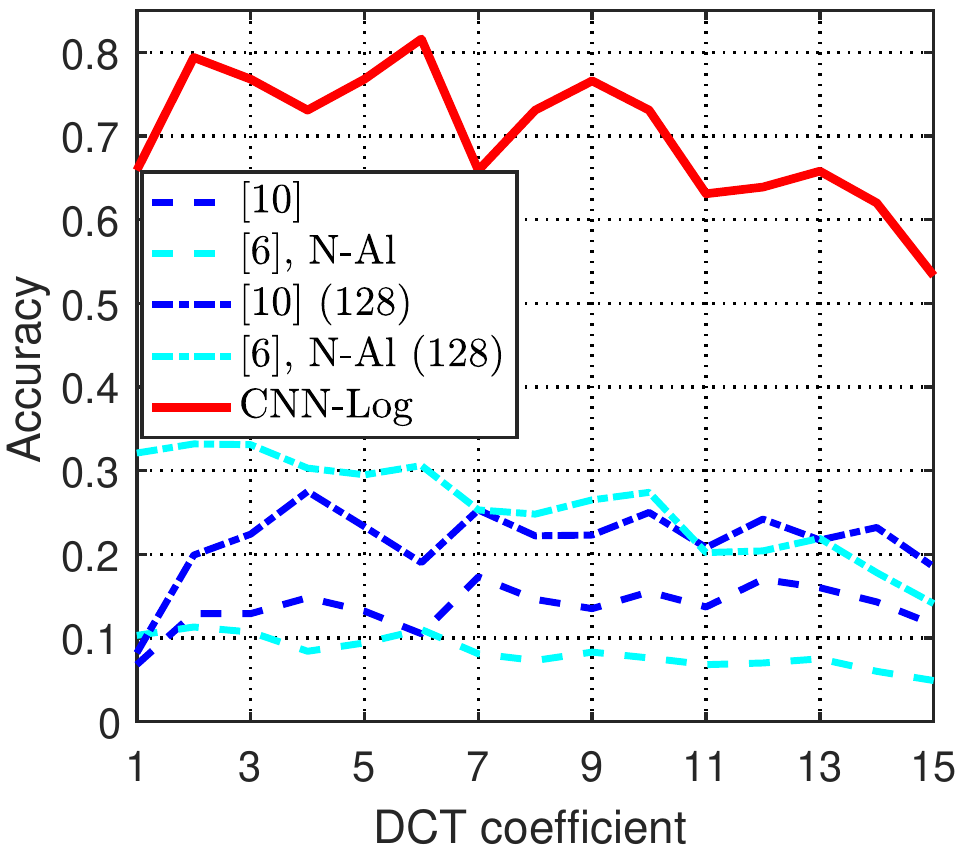}}
\subfigure{
\label{Fig1_2}
\includegraphics[width = 0.65\columnwidth, width = 0.48\columnwidth]{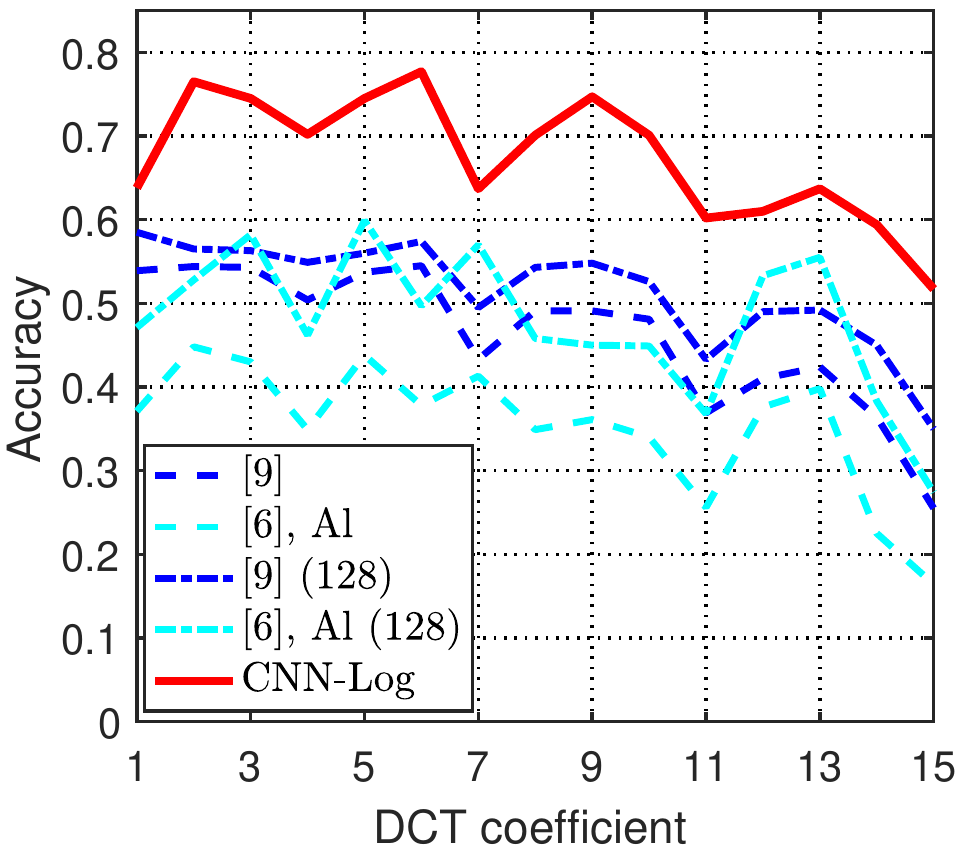}}
\caption{Average Acc of the estimation for each DCT coefficient in the non-aligned (left) and aligned (right) case ($QF_2 = 90$).}
\label{fig.average_plot90}
\end{figure}

To better highlight the improvement in terms of accuracy, for the sota methods we also report the results when a larger window size of $128\times 128$ is used. We see that, even in this unfavourable case, the CNN estimator always outperforms the other methods in terms of MSE, while in terms of Acc, the CNN estimator provides the best performance in the non-aligned case, and when $QF_1 > QF_2$ in the aligned case.
We also observe that, in contrast to our CNN estimator, the other methods tend to provide much better performance in terms of accuracy than MSE. This can be explained by the fact that these methods tend to produce precise estimations in most cases at the cost of very large errors when an incorrect estimation is made.
To see the dependency of the performance on the DCT coefficient, Fig. \ref{fig.average_plot90} reports the estimation accuracy for each quantization step, i.e. $q_{1,i}$ for $i = 1 \dots 15$, averaged on all the $QF_1$'s. The better accuracy of the proposed system is confirmed for all DCT coefficients.

%
\begin{table}[htbp] 
\scriptsize
\centering
\caption{ CNN estimation performance and comparison, for $QF_2 = 90$, with mismatched $QF_1$, on RAISE dataset.}
    \begin{tabular}{|p{0.35cm}<{\centering}|p{1cm}<{\centering}p{1cm}<{\centering}p{0.85cm}<{\centering}
 |p{1cm}<{\centering}p{0.85cm}<{\centering}p{0.85cm}<{\centering}|}
    \hline
    \multirow{2}{*}{$QF_{1}$}& \multicolumn{3}{c|}{Non-Aligned}& \multicolumn{3}{c|}{Aligned}  \\ \cline{2-7}
    & CNN-Log &  \text{ \cite{Bianchi2012}, N-Al } & \text{ \cite{Dalmia2018} } &   CNN-Log &  \text{ \cite{Bianchi2012}, Al  } & \text{ \cite{Galvan2014} } \\
    \hline
    63 &\textbf{1.67}/\textbf{0.48}& 27.8/0.14 & 15.9/0.32 & \textbf{1.57}/0.42& 13.5/\textbf{0.62}& 11.2/0.61\\
    \hline
   78 & \textbf{0.62}/\textbf{0.51}& 24.8/0.08& 24.8/0.10& \textbf{0.63}/0.53& 18.2/0.36& 2.15/\textbf{0.72}\\
    \hline
   93 & \textbf{0.99}/\textbf{0.39}& 60.8/0.08& 61.9/0.00& \textbf{1.42}/0.20& 44.1/\textbf{0.22}& 4.85/0.00\\
    \hline
    \end{tabular}%
\label{tab.Res90mismatchQF1}%
\end{table}%

Table \ref{tab.Res90mismatchQF1} shows the results for values of $QF_1$ other than those considered during training, still in the range $[60:98]$. By comparing these results with those in Table \ref{tab.Res90}, we see that the performance decrease a bit, especially the accuracy values, but not seriously so. For completeness, the results are reported also in the aligned case. The case of $QF_2$ mismatch is considered in Table \ref{tab.Res92MismatchedQF2}, where a CNN trained with $QF_2 = 90$ is tested on images for which $QF_2 = 92$. The performance drop is negligible, hence proving a certain generalization capability of the network.
\begin{table}[htbp] 
\scriptsize
\centering
\caption{ CNN estimation performance and comparison for mismatched $QF_2 = 92$, on RAISE dataset.}
   \begin{tabular}{|p{0.35cm}<{\centering}|p{1cm}<{\centering}p{1cm}<{\centering}p{0.85cm}<{\centering}
 |p{1cm}<{\centering}p{0.85cm}<{\centering}p{0.85cm}<{\centering}|}
    \hline
    \multirow{2}{*}{$QF_1$}& \multicolumn{3}{c|}{Non-Aligned}& \multicolumn{3}{c|}{Aligned}  \\ \cline{2-7}
    &CNN-Log &  \text{ \cite{Bianchi2012}, N-Al } & \text{ \cite{Dalmia2018} } &  CNN-Log  &  \text{ \cite{Bianchi2012}, Al} & \text{ \cite{Galvan2014} } \\
    \hline
    75&\textbf{0.51}/\textbf{0.63}&21.1/0.10& 18.5/0.18&\textbf{0.33}/\textbf{0.78}&14.3/0.52&3.39/0.72\\\hline
    80&\textbf{0.61}/\textbf{0.62}&25.2/0.08& 26.7/0.06&\textbf{0.28}/\textbf{0.83 }& 16.1/0.42& 1.96/0.77\\\hline
    85&\textbf{0.56}/\textbf{0.68} & 33.3/0.08 & 42.2/0.07 & \textbf{0.31}/\textbf{0.85} & 24.3/0.29 & 1.31/0.78 \\
    \hline
    \end{tabular}%
\label{tab.Res92MismatchedQF2}%
\end{table}
\begin{table}[t!] 
\scriptsize
\centering
\caption{Performance of the CNN estimator (MSE/Acc), for $QF_2 = 90$ on Dresden dataset.}
    \begin{tabular}{|p{0.35cm}<{\centering}|p{1cm}<{\centering}p{1cm}<{\centering}p{0.85cm}<{\centering}
 |p{1cm}<{\centering}p{0.85cm}<{\centering}p{0.85cm}<{\centering}|}
    \hline
    \multirow{2}{*}{$QF_1$}& \multicolumn{3}{c|}{Non-Aligned}& \multicolumn{3}{c|}{Aligned}  \\ \cline{2-7}
   & CNN-Log &  \text{ \cite{Bianchi2012}, N-Al } & \text{ \cite{Dalmia2018} } & CNN-Log &  \text{ \cite{Bianchi2012}, Al  } & \text{ \cite{Galvan2014} } \\
    \hline
  60&\textbf{0.71}/\textbf{0.61}&37.3/0.13&24.7/0.28&\textbf{0.61}/\textbf{0.65}&22.6/0.48&19.8/0.49\\\hline
  65&\textbf{0.61}/\textbf{0.58}&28.4/0.12&20.1/0.27&\textbf{0.63}/\textbf{0.58}&15.1/0.53&13.6/0.50\\ \hline
  70&\textbf{0.54}/\textbf{0.61}&22.6/0.11&19.1/0.22&\textbf{0.45}/\textbf{0.64}&9.77/0.50&7.67/0.55\\\hline
  75&\textbf{0.34}/\textbf{0.74}&22.4/0.10&20.8/0.18&\textbf{0.19}/\textbf{0.85}&14.3/0.45&4.87/0.60\\\hline
  80&\textbf{0.37}/\textbf{0.73}&27.1/0.07&28.1/0.07&\textbf{0.16}/\textbf{0.85}&13.7/0.34&2.35/0.65\\\hline
  85&\textbf{0.39}/\textbf{0.73}&36.3/0.07&31.7/0.13&\textbf{0.30}/\textbf{0.81}&30.3/0.23&2.31/0.78\\\hline
  90&\textbf{0.68}/\textbf{0.59}&50.1/0.10&44.2/0.00&2.23/\textbf{0.19}&38.9/0.00&\textbf{2.17}/0.00\\\hline
  95&\textbf{1.22}/\textbf{0.45}&69.0/0.05&61.3/0.00&\textbf{0.64}/\textbf{0.59}& 59.0/0.21&8.05/0.00\\\hline
  98&\textbf{1.22}/\textbf{0.56}&76.9/0.00&69.1/0.00&\textbf{1.21}/\textbf{0.55}&59.5/0.08&11.3/0.00\\
    \hline
    \end{tabular}%
\label{tab.Res90_Dresden}%
\end{table}%

In order to evaluate the impact of database mismatch, we carried out some tests on images belonging to the Dresden database. The results are reported in Table \ref{tab.Res90_Dresden}. Despite a small drop of the performance, the improvement with respect to the sota is still evident. We also run some tests on images for which the first compression is carried out by using Photoshop (the second compression is the same as before). These are particularly significant results since Photoshop does not use a standard quantization matrix, thus resulting in a strong mismatch between the training and test data. The results we got are reported in Table \ref{tab.Res90_PS} for some values of the quality measure adopted by Photoshop (PS). Upon inspection of the table, we see that the performance of the CNN estimator remain reasonably good in the non-aligned case, where they significantly outperform those obtained by sota methods, especially, but not only, in terms of MSE. In the aligned case, model-based methods work better for lower values of PS. The reason for such a behaviour is that model-based methods are less sensitive to the specific values of the quality factor than data-driven techniques. For large values of PS, the CNN-based estimator works much better than the sota, since in such cases the first quantization step is smaller than the second one, a situation that model-based techniques can not handle properly.

\begin{table}[htbp] 
\scriptsize
\centering
\caption{ CNN estimation performance and comparison when the first JPEG is done with Photoshop ($QF_2 = 90$), on RAISE dataset.}
   \begin{tabular}{|p{0.35cm}<{\centering}|p{1cm}<{\centering}p{1cm}<{\centering}p{0.85cm}<{\centering}
 |p{1cm}<{\centering}p{0.85cm}<{\centering}p{0.85cm}<{\centering}|}
    \hline
    \multirow{2}{*}{PS}& \multicolumn{3}{c|}{Non-Aligned}& \multicolumn{3}{c|}{Aligned}  \\ \cline{2-7}
    &CNN-Log &  \text{ \cite{Bianchi2012}, N-Al } & \text{ \cite{Dalmia2018} } &  CNN-Log  &  \text{ \cite{Bianchi2012}, Al} & \text{ \cite{Galvan2014} } \\
    \hline
    8&\textbf{6.86}/0.18&28.2/0.07& 23.5/\textbf{0.21}&6.49/0.18&16.4/0.43&\textbf{5.11}/\textbf{0.70}\\\hline
    9& \textbf{3.17}/\textbf{0.27}&32.8/0.07& 36.6/0.09& 4.81/0.22& 29.3/0.24& \textbf{1.98}/\textbf{0.75}\\\hline
    10 & \textbf{1.83}/\textbf{0.33} & 50.1/0.08 & 58.1/0.02 & 4.65/0.05 & 50.5/0.07 & \textbf{2.68}/\textbf{0.29 } \\
    \hline
    11 & \textbf{1.09}/\textbf{0.60} & 68.4/0.05 & 78.3/0.00 & \textbf{1.02}/\textbf{0.60} & 56.7/0.17 & 7.56/0.00 \\
    \hline
    12 & \textbf{1.08}/\textbf{0.76} & 76.8/0.00 & 88.8/0.00 & \textbf{1.03}/\textbf{0.76} & 72.3/0.08 & 11.9/0.00\\
    \hline
    \end{tabular}%
\label{tab.Res90_PS}%
\end{table}

As a last test, we repeated all the experiments by letting $QF_2 = 80$ and $QF_1$ as detailed in Section \ref{sec.method}\footnote{These results have been obtained by retraining the CNN with $QF_2 = 80$.}. Due to lack of space, we report only a subset of the results we got (see Table \ref{tab.Res80}). Even in this case, the advantage of the proposed method is evident, despite a general performance loss by all techniques, since many sota methods tend to perform worse when the second $QF$ is small (noticeably smaller that $QF_1$).
\begin{table}[htbp] 
\scriptsize
\centering
\caption{ Performance of the CNN estimator (MSE/Acc), for $QF_2 = 80$, on RAISE dataset, and comparison.}
    \begin{tabular}{|p{0.35cm}<{\centering}|p{1cm}<{\centering}p{1cm}<{\centering}p{0.85cm}<{\centering}
 |p{1cm}<{\centering}p{0.85cm}<{\centering}p{0.85cm}<{\centering}|}
    \hline
    \multirow{2}{*}{$QF_{1}$}& \multicolumn{3}{c|}{Non-Aligned}& \multicolumn{3}{c|}{Aligned}  \\ \cline{2-7}
    & CNN-Log &  \text{ \cite{Bianchi2012}, N-Al } & \text{ \cite{Dalmia2018} } &   CNN-Log &  \text{ \cite{Bianchi2012}, Al  } & \text{ \cite{Galvan2014} } \\
    \hline
    55&\textbf{2.61}/\textbf{0.51}&39.6/0.09& 40.7/0.05&\textbf{3.03}/0.26& 27.9/\textbf{0.47}& 12.9/0.43\\
    \hline
    60&\textbf{1.75}/\textbf{0.46}&27.7/0.08& 28.2/0.08& \textbf{1.28}/\textbf{0.56}& 20.8/0.37&7.28/0.43\\
    \hline
    65&\textbf{1.38}/\textbf{0.44}&22.1/0.07& 20.4/0.09& \textbf{1.36}/0.36& 18.2/0.32& 5.42/\textbf{0.49}\\
    \hline
    70&\textbf{1.20}/\textbf{0.48}&20.7/0.07 &16.5/0.11& \textbf{0.93}/0.54& 18.2/0.41& 3.19/\textbf{0.58}\\
    \hline
    75&\textbf{1.07}/\textbf{0.51}&23.8/0.06 &16.7/0.11& 10.6/0.20& 28.5/0.15 &\textbf{1.30}/\textbf{0.66}\\
    \hline
    80&\textbf{1.67}/\textbf{0.38}&32.7/0.07 &22.2/0.06& 14.4/\textbf{0.01}& 30.8/0.00 &\textbf{2.57}/0.00\\
    \hline
    85&\textbf{1.80}/\textbf{0.34}&46.0/0.05& 31.5/0.05& \textbf{5.67}/0.05& 29.4/\textbf{0.17}& 9.23/0.00\\
    \hline
    90&\textbf{1.86}/\textbf{0.40}&65.1/0.11& 46.4/0.00& \textbf{1.39}/\textbf{0.50}&55.6/0.06 &20.9/0.00\\
    \hline
    95 & \textbf{3.18}/\textbf{0.38}& 88.2/0.00& 66.1/0.00&\textbf{2.98}/\textbf{0.42}& 73.6/0.04 &37.2/0.00\\
    \hline
    \end{tabular}%
\label{tab.Res80}%
\end{table}%

\section{Concluding remarks}
\label{sec.conc}
We proposed a general method for primary quantization matrix estimation based on CNN, which can work under a wide variety of conditions. Since the operative conditions are usually unknown when the method is applied in practice, an algorithm that can work in all the conditions represent a great advantage with respect to the use of dedicated estimators.  Despite its generality, the proposed method outperforms the existing - dedicated - solutions in most of the cases. Two distinctive features of the new method are its capability to retain very good performance also in the challenging case of $QF_1 > QF_2$, and the good performance it achieves on small image patches. The latter characteristic is particularly important when the estimation of the quantization matrix is a preliminary step towards image tampering localization, since in this way the resolution of localization would improve.

Given the highly discretized nature of the quantization steps, future work could consider casting the estimation problem as a classification task: we expect that this may allow to reach a better accuracy, possibly at the expenses of a larger MSE.

\section*{Acknowledgements}
This work has been partially supported by a research sponsored by DARPA and Air Force Research Laboratory (AFRL) under agreement number FA8750-16-2-0173.




\clearpage

\bibliographystyle{IEEEtran}
\bibliography{SPLQ1matrix}

\end{document}